\documentclass{article}

\usepackage{proof}
\usepackage{amssymb}
\usepackage{amsmath}
\usepackage{fullpage}
\usepackage{lscape}
\usepackage{mathabx}

\title{Exposition: Synthesis via Functional Interpretation}
\author{Daniel Weller}
\date{March 2, 2014}

\newcommand{\impl}{\rightarrow}

\newcommand{\allelim}{\forall_e}
\newcommand{\allintro}{\forall_i}
\newcommand{\exelim}{\exists_e}
\newcommand{\exintro}{\exists_i}
\newcommand{\orelim}{\lor_e}
\newcommand{\orintro}{\lor_i}
\newcommand{\andintro}{\land_i}
\newcommand{\implelim}{\impl_e}
\newcommand{\indrule}{\mathrm{IND}}

\newcommand{\extractOp}{\mathcal{E}}
\newcommand{\extract}[1]{\extractOp(#1)}
\newcommand{\nocomp}{\varepsilon}

\newcommand{\nats}{\mathbb{N}}
\newcommand{\subst}[2]{[ #1 := #2]}

\newcommand{\ifthenelse}[3]{\mathbf{If}\, #1\, \mathbf{Then}\, #2\, \mathbf{Else}\, #3}
\newcommand{\ifthenelselines}[3]{
\begin{matrix}
\mathbf{If}\, #1\\
\mathbf{Then}\, #2\\
\mathbf{Else}\, #3
\end{matrix}}
\newcommand{\recOp}{\mathbf{R}}
\newcommand{\recursion}[3]{\recOp #1 #2 #3}
\newcommand{\reduces}{\rightarrow}
\newcommand{\cminus}{\dotdiv}
\newcommand{\isZero}{\mathbf{isZero}}
\newcommand{\pleft}{\mathbf{left}}
\newcommand{\pright}{\mathbf{right}}

\begin{document}
\maketitle
The aim of this short paper is to give a practical introduction to functional interpretation of proofs
in arithmetic for computer scientists interested in synthesis. Towards this, we will define our own notion
of functional interpretation which differs (sometimes only inessentially) from those used in the literature,
but has the advantage (in the
opinion of the author) of being very natural. Note that we only show functional interpretation
by definition and example --- it is quite possible that proving the correctness of this formalism
is cumbersome (or even impossible). Still, as can be witnessed below, our formalism allows extraction
of a correct program from a non-trivial proof in a systematic way, hopefully elucidating a few
central ideas common to all notions of functional interpretation.

With this disclaimer in mind, we can start setting up our machinery for functional 
interpretation.
Towards extracting a program from a proof, we have to fix (1) a programming language, (2)
a proof system, and (3) a translation from (2) to (1).

\textbf{The programming language.} Choosing a suitable programming language for functional extraction 
involves some design decisions. Usually, one takes a functional programming language, since this
often induces a simple and natural translation from proofs to programs. Some other decisions are more inessential,
e.g.~how to represent boolean values in the programming language (e.g.~by constants $\bot, \top$, or by numerals
$0,1$).

For simplicity, we choose
the untyped $\lambda$-calculus\footnote{All the $\lambda$-terms obtained by functional extraction
are actually typable. We chose to use the untyped $\lambda$-calculus since in the context of this
exposition, we view types as a distraction. Most works on functional extraction {\em do} work with
a typed $\lambda$-calculus, since it is useful in e.g.~proving correctness.}
extended with some constants for arithmetic, pairs, and program control. More precisely, we assume existence
of a countable set of variables $V$, and a fixed set of constant symbols
$\{0,1,+,\cminus,\mathbf{pair}, \pleft, \pright, \isZero, \mathbf{If Then Else}, \mathbf{R}\}$.
Variables and constants are {\em $\lambda$-terms},
and if $s,t$ are $\lambda$-terms and $x$ a variable then $st$ and $\lambda x.t$ are $\lambda$-terms.
We use infix notation for $+,\cminus$ and write
$\mathbf{If Then Else}t_1t_2t_3$ as $\ifthenelse{t_1}{t_2}{t_3}$
and $\mathbf{pair}t_1t_2$ as $(t_1, t_2)$). The intended semantics
of the symbols are clear except maybe for $\mathbf{R}$, which will be the recursion operator.
Variable-free terms consisting only of $0,1,+$ may denote numbers (i.e.~$1+(1+1)$ and $(1+1)+1$ both
denote $3\in\nats$, while $+++$ does not denote a number). Such terms are called {\em arithmetical},
and we will often not distinguish between a number $\alpha\in\nats$ and the arithmetical terms that
denote it. In particular, if $\alpha\in\nats$ and $t$ is a $\lambda$-term, then by $t\alpha$ we denote
the $\lambda$-term which applies $t$ to the numeral representing $\alpha$.

The formal semantics of our programming language are given by the reduction rules:
\begin{align*}
  (\lambda x.t)s &\reduces t\subst{x}{s},\\
  \pleft (s, t)& \reduces s,\\
  \pright (s, t)& \reduces t,\\
  \isZero(0)&\reduces 0,\\
  \isZero(t+1)&\reduces 1,\\
  \ifthenelse{0}{t}{s}& \reduces t,\\
  \ifthenelse{1}{t}{s}& \reduces s,\\
  \recursion{b}{s}{0}&\reduces b,\\
  \recursion{b}{s}{(t+1)}&\reduces st(\recursion{b}{s}{t}),\\
  t\cminus s&\reduces u\quad\mbox{if } t,s \mbox{ arithmetical denoting }a,b,\mbox{ and } a>b,\mbox{ and } u\mbox{ denotes }a-b,\\
  t\cminus s&\reduces 0\quad\mbox{if } t,s \mbox{ arithmetical denoting }a,b,\mbox{ and } a\leq b,
\end{align*}
where $\subst{x}{s}$ denotes capture-avoiding substitution. 
Hence we have $\beta$-reduction and the usual defining reductions for our constant symbols, where
we have chosen to represent ``true'' by 0 and ``false'' by 1.
Note that, in the clauses for $\recOp$, the term $b$ corresponds to the 
base case of a recursive definition, the term $s$ corresponds to the step case, and $s$ will usually
be of the form $\lambda x\lambda y.t(x,y)$, where the variable $x$ corresponds to the recursion counter
and the variable $y$ to the result of the recursive call. $\cminus$ denotes the usual ,,cutoff subtraction''
on the natural numbers.

It is fair to call this system a programming language: the set of terms is recursive and the relation
$\reduces$ has low computational complexity. It is easy to see that an interpreter for this language 
(i.e.~an implementation of the transitive, reflexive, compatible closure of the $\reduces$ relation)
can be written in any Turing-complete programming language.

\textbf{Proof system.} We use natural deduction for intuitionistic logic
with equality and induction (over the language $\{0,1,+,=,\geq\}$). For the sake of conciseness, we only present
the subset of rules that we will use in the example presented later in this paper.
\[
  \infer[\andintro]{A\land B}{A & B}
\qquad
  \infer[\orelim]{C}{A\lor B & \deduce{C}{\deduce{\vdots}{[A]}}
			     &\deduce{C}{\deduce{\vdots}{[B]}}}
\qquad
\infer[\orintro]{A\lor B}{A}
\qquad
\infer[\orintro]{A\lor B}{B}
\qquad
\infer[\implelim]{B}{A & A\impl B}
\]
\[
\infer[\allintro]{\forall x.A}{A}
\qquad
\infer[\allelim]{A\subst{x}{t}}{\forall x.A}
\qquad
\infer[\exelim]{A}{\exists x.B & \deduce{A}{\deduce{\vdots}{[B]}}}
\qquad
\infer[\exintro]{\exists x.A}{A\subst{x}{t}}
\]
\[
\infer[\indrule]{\forall x.A}{A(0) & \deduce{A(x+1)}{\deduce{\vdots}{[A(x)]}}}
\qquad
\infer[=]{A(t)}{t=s & A(s)}
\]
where, as usual, $[A]$ denotes discharging an assumption $A$ and the $\exelim,\allintro$
rules have an eigenvariable condition.
At the leaves of trees constructed by these
rules, we allow only discharged assumptions and (non-discharged) {\em axioms}, which we
take to be $t=t$, $t\geq t$,
$t=0 \lor \exists y. t=y+1$, $t \geq s \impl t+1 \geq s+1$,
and $t\geq 0$ for all terms $s,t$.

Note that while our proof system is not directly suitable for automated proof search (in contrast
to e.g.~resolution proof systems), most other proof systems can be polynomially translated into our
system.

\textbf{Program extraction.}
We will now define a map $\extractOp$ from proofs to $\lambda$-terms with the intention
that for a proof $\pi$ of $\forall x\exists y. F(x,y)$ we will have\footnote{This is not precisely
true: actually, $\extract{\pi}(n)$ will be a pair s.t.~$\nats\models F(n,\pleft(\extract{\pi}(n)))$.}
$\nats\models F(n,\extract{\pi}(n))$ for all $n\in\nats$, where $\models$
is the usual semantic consequence operator (which in particular interprets
our $\lambda$-terms as functions in the natural way). 
In other words, $\extract{\pi}$, when 
viewed as function $\nats\to\nats$, fulfills the specification $F$.

The idea in defining $\extractOp$ will be to construct the ``computational content''
of a proof from the computational content of its premises. The most basic idea is that
when we have a proof of $B$ from an assumption $[A]$, then the computational content
of $B$ will depend upon that of $A$. In our setting, this means that
the assumption $[A]$ induces a variable $x_A$ in the computational content of $B$,
and this variable will at some
point be substituted by some other computational content as determined by the proof.
In general, the type of the computational content will be closely related to the formula
that is derived; for example a proof of  $\exists x.F$ will have as computational content
a pair $(t,c)$ where $t$ is the witness of $\exists x$, and $c$ is the computational
content of the proof of $F\subst{x}{t}$. Note that, unsurprisingly,
the computational interpretations of $\exists$ and $\lor$ are closely related,
as are the interpretations of $\forall$ and $\land$. Roughly speaking, the propositional
structure of the proof determines the structure of {\em functionals and control} in the program,
while the quantifiers determine the structure of {\em data} in the program.

It is immediately clear that some proofs do
not contain computational content (e.g.~a proof of $A\impl A$)\footnote{Note that this is an important
distinction between the interpretation of proofs as functions in our setting, and the setting
of the Curry-Howard isomorphism as it is classically understood: in that setting, 
the proofs of $A\impl A$ are exactly
the programs $\nats\to\nats$.}, therefore it will be useful
to fix a variable $\nocomp$ which we will use to denote ``no computational content''.
\footnote{In the typed setting, one would introduce an accompanying type $\nocomp$ for
``no computational content'', and one would propagate this information as much as
possible. E.g.~one would identify the types $A\to\nocomp$ and $\nocomp$, since $A\to\nocomp$
would be the type of a function taking an object of type $A$, and returning something
which does not have computational content. Functions of such types may appear when functional
extraction is done naively, and one wants to avoid creating $\lambda$-terms of such types
for efficiency reasons.}

The map $\extractOp$ is defined by structural induction on natural deduction proofs.
For discharged assumptions, we set
$\extract{[A]}:= x_A$ (i.e.~all discharged assumptions $A$ are assigned the same
variable $x_A$).
For the axioms, we mostly assign no computational content by setting
$\extract{t=t}:=\nocomp$, $\extract{t\geq t}:=\nocomp$ and
$\extract{t \geq s \impl t+1 \geq s+1}:=\lambda x.\nocomp$, except that
we set
\[
  \extract{t=0 \lor \exists y. t=y+1}:=(\isZero(t),\ifthenelse{\isZero(t)}{\nocomp}{(t\cminus 1, \nocomp)}).\footnote{As expected, the computational content intuitively corresponds to the witness of the $\exists y$
quantifier (structured in the correct way --- compare this with the definitions in Table~\ref{tab:def:extract}). But note that $\cminus$ is not part of the language used in our proof
system: it is only contained in
our programming language. Still, we have $\nats \models t=0 \lor t=(t\cminus1)+1$ as
expected. This is an example of the general observation that all axioms which have
a computational interpretation in the programming language can be added to the proof system.}
\]

The definition of $\extractOp$ by case distinction on the last rule in a proof $\pi$ can be found
in Table~\ref{tab:def:extract}. The intuition behind the definition is the following: a proof of a conjunction
contains computational content for both its subproofs, a proof of a disjunction contains information
which disjunct is true, and the computational content of that disjunct, a proof that eliminates a disjunction
corresponds to a case distinction on whether the left or right conjunct is true (and passes on the computational
content of the proof of this disjunct), and so on.

Instead of going the usual way of proving correctness of the translation, we will instead apply these definitions
to an example and verify that indeed, witness-computing programs are extracted.

\begin{table}
  \begin{center}
\begin{tabular}{|c|c|}
  \hline
  $\pi$ & $\extract{\pi}$ \\
  \hline
  $\infer[\andintro]{A\land B}{\deduce{A}{(\pi_1)} & \deduce{B}{(\pi_2)}}$ & $(\extract{\pi_1}, \extract{\pi_2})$\\
  \hline
  $\infer[\orintro]{A\lor B}{\deduce{A}{(\pi_1)}}$ & $(0, \extract{\pi_1})$\\
  \hline
  $\infer[\orintro]{A\lor B}{\deduce{B}{(\pi_1)}}$ & $(1, \extract{\pi_1})$\\
  \hline
  $\infer[\orelim]{C}{\deduce{A\lor B}{(\pi_1)} & \deduce{C}{\deduce{\vdots (\pi_2)}{[A]}}
																						     &\deduce{C}{\deduce{\vdots (\pi_3)}{[B]}}}$ & $\ifthenelselines{\pleft(\extract{\pi_1})}{\extract{\pi_2}\subst{x_A}{\pright(\extract{\pi_1})}}{\extract{\pi_3}\subst{x_B}{\pright(\extract{\pi_1})}}$\\
  \hline
  $\infer[\implelim]{B}{\deduce{A}{(\pi_1)} & \deduce{A\impl B}{(\pi_2)}}$ & $\extract{\pi_2}\extract{\pi_1}$\\
  \hline
  $\infer[\allintro]{\forall x.A}{\deduce{A}{(\pi_1)}}$ & $\lambda x.\extract{\pi_1}$ \\
  \hline
  $\infer[\allelim]{A\subst{x}{t}}{\deduce{\forall x.A}{(\pi_1)}}$ & $\extract{\pi_1}t$ \\
  \hline
  $\infer[\exelim]{A}{\deduce{\exists x.B}{(\pi_1)} & \deduce{A}{\deduce{\vdots(\pi_2)}{[B]}}}$
											      & 
  $\extract{\pi_2}\subst{x}{\pleft(\extract{\pi_1})}\subst{x_B}{\pright(\extract{\pi_1})}$\\
\hline
$\infer[\exintro]{\exists x.A}{\deduce{A\subst{x}{t}}{(\pi_1)}}$ & $(t, \extract{\pi_1})$\\
\hline
$\infer[\indrule]{\forall n.A}{\deduce{A(0)}{(\pi_1)} & \deduce{A(n+1)}{\deduce{\vdots (\pi_2)}{[A(n)]}}}$ &
$\lambda u.\recursion{(\extract{\pi_1})}{(\lambda n\lambda x_{A(n)}.\extract{\pi_2})u}$\\
\hline
$\infer[=]{A(t)}{\deduce{t=s}{(\pi_1)} & \deduce{A(s)}{(\pi_2)}}$ & $\extract{\pi_2}$\\
\hline
\end{tabular}
\caption{Extraction of computational content from proofs.}\label{tab:def:extract}
\end{center}
\end{table}

\textbf{Example.} 
We show how to synthesize the maximum function $\max:\nats\times\nats\to \nats$
from its specification
in our setting.
Letting 
\[
  F(x_1,x_2,y)= y\geq x_1 \land y\geq x_2 \land (y=x_1\lor y=x_2),
\]
the $\max$ function has the specification $\forall x_1,x_2\in\nats: F(x_1,x_2,\max(x_1,x_2))$.
We extract a functional program realizing $\max$ from a natural deduction proof 
of $\forall x_1,x_2\exists y.F(x_1,x_2,y)$
using the lemma $L:=\forall x_1,x_2.x_1\geq x_2 \lor x_2\geq x_1$.
Let $\pi$ be the proof
\small
\[
  \infer[\allintro]{\forall x_1,x_2\exists y.F(x_1,x_2,y)}{
    \infer[\orelim]{\exists y.F(x_1,x_2,y)}{
      \infer[\allelim]{x_1\geq x_2 \lor x_2\geq x_1}{
      \deduce{\forall x_1,x_2.x_1\geq x_2 \lor x_2\geq x_1}{(\psi)}}
	&
	\infer[\exintro]{\exists y.F(x_1,x_2,y)}{
	  \infer[\andintro]{F(x_1,x_2,x_1)}{x_1 \geq x_1 & [x_1 \geq x_2] &
      \infer[\orintro]{x_1 = x_1 \lor x_1=x_2}{x_1=x_1}}}
	&
	\infer[\exintro]{\exists y.F(x_1,x_2,y)}{
	  \infer[\andintro]{F(x_1,x_2,x_2)}{[x_2 \geq x_1] & x_2 \geq x_2 &
      \infer[\orintro]{x_1 = x_2 \lor x_2=x_2}{x_2=x_2}}}
    }
}
\]
\normalsize
For the moment, we omit how exactly the proof $(\psi)$ of the lemma 
$L$
is treated.
We construct the computational content of $\pi$ according to 
the interpretation of the leafs and Table~\ref{tab:def:extract}. 
Letting $f$ denote the computational
content of $\psi$, and letting (for easier readability) $x_{x_1\geq x_2}=Y$ and $x_{x_2\geq x_1}=Z$, we obtain the following.
The left $\exintro$ induces the $\lambda$-term $(x_1,(\nocomp,(Y,(0,\nocomp))))$ and the
right $\exintro$ induces the $\lambda$-term $(x_2,(Z,(\nocomp,(1,\nocomp))))$. Putting things
together, we obtain for $\pi$ the $\lambda$-term $\extract{\pi}$:
\[
  \lambda x_1x_2.\ifthenelse{\pleft(f x_1 x_2)}
  {(x_1,(\nocomp,(\pright(f x_1x_2),(0,\nocomp))))}
  {(x_2,(\pright(f x_1x_2),(\nocomp,(1,\nocomp))))}.
\]
Now let $\alpha_1,\alpha_2$ be numerals. 
Assuming that $f$ is interpreted correctly (i.e.~that
$\pleft(f\alpha_1\alpha_2)$ normalizes to $0$ if $\alpha_1\geq \alpha_2$ and $1$ otherwise), the term for $\pi$,
when applied to $\alpha_1,\alpha_2$, normalizes correctly either to $(\alpha_1,\ldots)$ or $(\alpha_2,\ldots)$,
where $\ldots$ contains the computational content of the conjuncts of $F(\alpha_1,\alpha_2,\alpha_i)$ 
(which in this case is anyways empty since $F(x,y,z)$ is quantifier-free). 
Hence $\pleft(\extract{\pi}\alpha_1\alpha_2)$ computes $\max(\alpha_1,\alpha_2)$ as
desired.
%

Regarding the proof $\psi$ of $L$, we have two options: either we assume that we 
have a program that, given $\alpha_1,\alpha_2\in\nats$ decides whether $\alpha_1\geq \alpha_2 \lor \alpha_2\geq \alpha_1$
(in this case, we treat $L$ as an axiom),
or we prove $L$ and synthesize the program
from the proof. In practice, the first option is more reasonable, but for sake of
exposition we take the second option here: indeed, the proof of the lemma involves induction
and therefore gives rise to a recursive program.
Setting $A(x_1):=\forall x_2.x_1\geq x_2 \lor x_2\geq x_1$, we let $\psi$ be
\[
  \infer[\indrule]{\forall x_1 A(x_1)}{
    (\psi_{\mathrm{b}})
    &
    (\psi_{\mathrm{s}})
  }
\]
where $\psi_{\mathrm{b}}$ is 
\[
\infer[\allintro]{A(0)}{\infer[\orintro]{0\geq x_2 \lor x_2 \geq 0}{x_2 \geq 0}}
\]
and $\psi_{\mathrm{s}}$ is 
\[
\infer[\allintro]{A(n+1)}{\infer[\orelim]{n+1\geq x_2 \lor x_2\geq n+1}{
	x_2=0 \lor \exists y. x_2=y+1 &
    (\varphi_{\mathrm{l}})
      &
    (\varphi_{\mathrm{r}})
    }
      }
\]
where $\varphi_{\mathrm{l}}$ is
\[
  \infer[=]{n+1\geq x_2 \lor x_2\geq n+1}{[x_2=0] & \infer[\orintro]{n+1\geq 0 \lor x_2\geq n+1}{n+1\geq 0}}
\]
and $\varphi_{\mathrm{r}}$ is
\[
\infer[\exelim]{n+1\geq x_2 \lor x_2\geq n+1}{
  [ \exists y. x_2=y+1 ]
    &
  \infer[=]{n+1\geq x_2 \lor x_2\geq n+1}{ [x_2 = y+1] &
  (\varphi)
  }}
\]
where $\varphi$ is
\[
\infer[\orelim]{n+1\geq y+1 \lor y+1\geq n+1}{
	  \infer[\allelim]{n\geq y \lor y \geq n}{[A(n)]} &
	  \infer[\orintro]{n+1\geq y+1 \lor y+1 \geq n+1}{\infer[\implelim]{n+1\geq y+1}{ [n \geq y] & n \geq y \impl n+1\geq y+1}} &
          \infer[\orintro]{n+1\geq y+1 \lor y+1 \geq n+1}{\infer[\implelim]{y+1\geq n+1}{ [y \geq n] & y \geq n \impl y+1\geq n+1}}
      }
\]
Setting $x_{A(n)}=Z$ and $x_{x_2=y+1}=U$ for readability, functional extraction yields $\extract{\psi}$ (after application of some
reduction rules to improve readability):
\[
  \underbrace{\lambda u.\recursion{\underbrace{(\lambda x_2.(1,\nocomp))}_{\extract{\psi_\mathrm{b}}}}{(\lambda n\lambda Z\underbrace{\lambda x_2.
	\ifthenelse{\isZero(x_2)}{\underbrace{(0,\nocomp)}_{\extract{\varphi_{\mathrm{l}}}}}{\underbrace{\ifthenelse{\pleft(Z(x_2\cminus 1))}{(0,\nocomp)}
	{(1,\nocomp)}}_{\extract{\varphi_{\mathrm{r}}}=\extract{\varphi}\subst{y}{x_2\cminus 1}\subst{U}{\nocomp} }}
  )}_{\extract{\varphi_{\mathrm{s}}}}}{u}}_{\extract{\psi}}
\]
One can check that for all $\alpha_1,\alpha_2\in\nats$, we have that if $\extract{\psi}\alpha_1\alpha_2$ 
reduces to $(0,\ldots)$, then $\alpha_1\geq\alpha_2$, and if it reduces to $(1,\ldots)$, then 
$\alpha_2\geq\alpha_1$, and that this term always reduces to one of these two forms.
Note that the length of the reduction sequence is linear in $\alpha_1$ since we recurse 
from $\alpha_1+1$ to $\alpha_1$. A logarithmic algorithm 
(corresponding to the comparison of the binary representations of $\alpha_1,\alpha_2$) could be obtained by using
``binary induction'' $A(0)\land (\forall x.A(x)\impl A(0x)\land A(1x)) \impl \forall x A(x)$.

\textbf{Soundness.} For our purposes, the most important notion of soundness is that from proofs
of $\Pi_2$-statements, i.e.~statements of the form $\forall x\exists y.F(x,y)$, with $F(x,y)$ quantifier-free, we can extract
programs that compute a correct $y$ given an $x$, as indicated above. To do this, one would define a binary
relation ``$t$ realizes $F$'', where $t$ is a $\lambda$-term and $F$ is a formula, by structural induction on $F$.
In particular, the definition would ensure that if $F$ is a $\Pi_2$-statement and $t$ realizes $F$, then $t$ is
a suitable program. One would finally show, by induction on natural deduction proofs $\pi$, that indeed $\extract{\pi}$
realizes $F$, where $F$ is the formula that $\pi$ proves. We refer to the literature for more details.

\textbf{Outlook.} There are many directions one can go from here. Note that we have only treated intuitionistic arithmetic ---
classical arithmetic can be treated by embedding it into intuitionistic logic (using e.g.~a double-negation translation etc.),
or directly by interpreting classical proofs or the law of excluded middle. We have not even given a computational interpretation
for all the usual rules of intuitionistic natural deduction (only what we used in our example proof) --- one could interpret
the whole system. Alternatively, one could just show how to interpret minimal logic (where the only connective is $\impl$),
and embed intuitionistic into minimal logic. One can investigate how the translation can be improved by removing redundant
parts of the extracted program (this can prevent construction of a term that is never used computationally in the proof).

\bibliographystyle{plain}
\bibliography{literatur}
\end{document}